\input mates.sty
\def\hb{\hfill\break}
\rightline{FTUAM 96-19}
\rightline{April, 1966}
\vskip.2cm
\centerline{ON THE THEORETICAL ANALYSIS OF THE SMALL $x$ 
DATA IN RECENT HERA PAPERS}
\vskip1.5cm
\setbox0=\vbox{\hsize65mm {\noindent\bf F. J. Yndur\'ain} 
\vskip .25cm
\noindent{\sl Departamento de F\'\i sica Te\'orica, C-XI,}\hb
{\sl Universidad Aut\'onoma de Madrid},\hb
Canto Blanco, \hb
E-28049 Madrid, SPAIN.\hb e-mail: fjy@delta.ft.uam.es}
\centerline{\box0}
\vskip1cm
{\bf Abstract}
\vskip.1cm
\setbox1=\vbox{\hsize12cm A number of internal inconsistencies,
 misleading conclusions and lack of completeness in 
the analyses of the small $x$ deep inelastic data in recent papers of 
HERA physicists, particularly in a very recent one by the H1 collaboration are 
pointed out. It is also shown that, when an analysis without prejudice is carried 
over, the situation is very different from the one claimed 
there in what respects checks of theoretical expectations.}
\centerline{\box1}
\vskip1cm
The so-called H1 Collaboration has recently
 produced a preprint concerning 
deep inelastic data at small $x$, that we will henceforth 
call HERA2. The work constitutes an improvement of the 
former HERA analysis$^{[1]}$ of the structure 
function $F_2$, to be denoted by HERA1, extending it to wider $x,\,Q^2$ ranges 
and greatly diminishing the errors. 

There is not much to be argued about the {\it experimental}
 obtaining of the data. The machine and 
the detectors worked wonderfully. The quality of the data is excellent. It 
is therefore more of a pity that the physicists in the 
collaboration have not been more careful with 
the subsequent theoretical analysis which shows 
 features that, to put it as mildly as possible, are difficult to justify.
 Because the second paper HERA2 (ref.2) is so much superior in the quality of 
the data, I will mostly discuss it here. Regarding it, I will address myself to a number 
of separate (but related) questions which I will 
now ennumerate: {\bf 1}) {\sl Parametrizations of the data}. {\bf 2}) {\sl 
Behaviour of structure functions as $x\rightarrow 0$}. {\bf 3}) {\sl Theoretical analysis with so-called 
``Double scaling limit"}. {\bf 4}) {\sl Connection with various theoretical 
models}. This is the central part of the present note, which will then be finished with some comments. 
\vskip.2cm
{\bf 1}.-The theory of strong interactions is QCD. On this
 practically everybody is agreed, so, when presenting parametrizations of 
data one should try to, at least, not violate grossly standard QCD features. In 
this, I think, it is no excuse that the parametrization is merely ``a 
phenomenological ansatz", as is described 
the parametrization of p. 20 of HERA2, Eq. (8). As should be by now widely 
known, the structure function $F_2$ may be split in a singlet and 
a nonsinglet piece, $F_2=F_S+F_{NS}$. For small $x$,
 $F_{NS}\simeq B_{NS} x^{0.5}$. {\it No matter what}, the fact that both 
$F_S$ and $F_{NS}$ are proportional to cross sections implies that $B_{NS}>0$: 
 ``phenomenological ansatz" or not. A measure of the quality of 
the fit is obtained by noting that the corresponding coefficient in (8), 
denoted by $e$ in the table underneath this Eq. (8) in HERA2 is {\it negative}. Really. 

But there is more. It may perhaps be argued that Eq.(8) was not meant 
to have anything to do with a QCD description: at least it 
should be compatible with the rest of the  paper HERA2. In the same page 20, 
the authors state that the wide range of $Q^2$ covered 
by their experiment allows them to study the behaviour at small 
$x$, and its variation with $Q^2$. So they write
$$F_2(x,Q^2)\simeqsub_{x\rightarrow 0}x^{-\lambda}\eqno (1)$$
and adscribe this prediction to De R\'ujula et al.$^{[3]}$ We will come 
to this last point later; for now what interests us is that the HERA2 collaboration 
find, and report in Table 4, values of $\lambda$ increasing as\footnote*{
We only report the values from $Q^2=12\;\gev^2$; below this 
value the 
contamination of nonsinglet is very strong, and even rough analyses 
shoud incorporate this fact to get any credibility.}
$$\lambda=0.24\;(Q^2=12\;\gev^2)\;
{\rm to}\;\lambda=0.50\;(Q^2=800\;\gev^2).\eqno (2)$$
Actually, a careful look at Table 4 (p.23) shows that, at least up to 
 $Q^2=350\;\gev^2$, $\lambda$ may be considered constant 
modulo what may easily be interpreted as statistical fluctuations, with 
a value of $\lambda$ around 
$\lambda\sim 0.32$. Again, we will return to this 
later; for now what interests me is the blatant contradiction 
between Eq.(2) and the parametrization of Eq.(8) where, for 
$x\rightarrow 0$, one has $F_2\simeq c\,x^d$ with $d=-0.188$, 
widely off the bounds given in Eq.(2). How the authors of HERA2 may claim that 
their data are fitted, {\it at the same time} by $x^{-0.32}$ and $x^{-0.19}$ 
baffles me. But there is more.
\vskip.2cm
{\bf 2}.-It is extremely important, for judging the quality of
 the theoretical analysis of HERA2, 
 to return to the matter of behaviour at 
small $x$. The authors of HERA2 fit $F_2$ at small $x$; and they
 are so certain of their results that they display them not only in the text, but in 
Table 4 (to which I have already referred). The {\it smallest} value of 
$\lambda$ they find is for $Q^2=2.5\;\gev^2$, where 
they get $\lambda=0.19$; for $Q^2=1.5\;\gev^2$ the value is $0.21$. In fact, 
an analysis at $Q^2=0$ (Compton scattering) gives$^{[4]}$ a $\lambda$ around 
0.26.

Then the authors of HERA2 state that ``the rate of growth of $F_2$ is expected 
to increase", and promptly signal to ref.3 as the place where such growth 
was predicted. So one opens ref. 3, and there one reads the prediction (p. 1651, 
line 9ff): ``The rate of growth increases with increasing $Q^2$. {\it It is always 
weaker than a power}" of $x$, although stronger than a log. (Italics mine).
 So, the HERA people who both in their phenomenological
 parametrization [Eq. (8) of HERA2] and in their Table 4 find {\it a power}, 
conclude that they check the prediction of ref. 3 --a surprising conclusion.
\vskip.2cm
{\bf 3}.-To clarify matters a bit more, I would 
like to bring attention to the derivation of the results of 
ref.3. Specifically, in p.1650 these authors consider the moments of the 
structure functions. They realize [their Eq. (2)] that 
the moments are given in terms of the Wilson expansion which, after a 
slight change of their rather old fashioned notation to the one prevalent 
nowadays, implies the relation
$$\mu_n(Q^2)\simeq \left[\frac{\alpha_s(Q^2_0)}{\alpha_s(Q^2)}\right]^{D(n)}a_n\simeq
\left[\frac{\log Q^2}{\log Q^2_0}\right]^{D(n)}a_n.\eqno (3)$$
The $a_n$ may easily be identified with expectation values of operators,
$$a_n\sim\langle p|O_n|p\rangle$$
that need not be specified here. The $D(n)$ are related to the anomalous dimensions. 
De R\'ujula et al. recognize that the behaviour of $F_2$ for small $x$ is linked 
to the singularities (in $n$, considered as a continuous variable)
 of the moments; the leading behaviour 
being given by the {\it rightmost} singularity. This is 
clear since one has 
$$\mu_n(Q^2)=\int^1_0\dd x\,F_2(x,Q^2)x^{n-2}.$$
 Now, if one {\it assumes} 
that at a certain $Q_0^2$ one has a Pomeron-type behaviour,
$$F_2(x,Q_0^2)\simeqsub_{x\rightarrow 0}C_0, C_0={\rm constant},\eqno (4)$$
then one has, at that $Q_0^2$, a singularity for $n=1$ that one can identify 
with that of the anomalous dimension, $D(n)$, which indeed is singular at $n=1$.

It should be emphasized that here one has a dichotomy. If at a given 
$Q_0^2$ one has a behaviour like a constant, then the structure function, by 
virtue of the analysis of De R\'ujula et al.$^{[3]}$ will grow ``always 
weaker than a power" of $x$, in their own words. Contrarywise, if at {\it any} $Q^2$, 
$F_2(x,Q^2)$ grows like a power, the conditions of the theorem cannot apply. 
Actually, the situation is really very simple. As discovered
 long ago$^{[6]}$, one has {\it two} possibilities. Either the singularities 
of the matrix elements $a_n$ lie to the left of those of $D(n)$, i.e., to the left 
of $n=1$, or to the right of it. If the first,
 one has the situation envisaged in ref.3. If, however, the dominating singularity
 of $a_n$ lies {\it to the right} of $n=1$, at a certain $n=1+\lambda,\,\lambda>0$,
 then one has a behaviour 
like a {\it power}, for all $Q^2$, $F_2(x,Q^2)\simeq x^{-\lambda}$. The 
coefficient of proportionality, a function of $Q^2$, may in fact be 
calculated. One may describe this result 
by stating that in the low $x$ regime, the dependence on $x$ and $Q^2$ of 
structure functions ``factorizes".

 The paper of De R\'ujula et al. is not very explicit. To find 
a more detailed calculation, we will refer to the review of 
Ball and Forte$^{[5]}$, apparently known to the HERA2 collaboration 
since they quote it. There, in p.9, the authors explicitely state that,
if $F_2\sim x^{-\lambda}$, the chain of 
arguments which lead to a ``double scaling form", namely 
their Eq.(2.31), breaks down, and a power-like behaviour becomes 
dominant [their Eq.(2.34), when one undoes their 
changes of variable, reads exactly $F_2(x,Q^2)\simeq x^{-\lambda}$]. Curiously enough,
  this power behaviour is exactly what is found by both HERA collaborations.

We can be more specific. Assume, with De R\'ujula et al,
 and Ball and Forte, that, contrary to the findings of the HERA2 group, 
one had,  at a given $Q^2_0$, a Pomeron-like behaviour as that given by Eq.(4) in the 
present paper. 
Then, as correctly stated in refs.3, 5, one obtains a behaviour 
dominated by the (known) singularities of $D(n)$. A simple 
calculation then gives, for all $Q^2\gg Q^2_0,\,x\rightarrow 0$, the 
very explicit formula
$$F_2(x,Q^2)\simeqsub_{x\rightarrow 0\atop Q^2\rightarrow \infty}
 C_0\left[ \frac{33-2n_f}{576\pi^2|\log x| \log[\alpha_s(Q_0^2)/\alpha_s (Q^2)]}
\right]^{\frac{1}{4}}
\exp \sqrt{\frac{144|\log x|}{(33-2n_f)}\;\left[\log\frac{\alpha_s(Q_0^2)}{\alpha_s(Q^2)}\right]}.
\eqno (5)$$
This equation is essentially identical
 to the corresponding one in ref.5, p.9; the  
behaviour it implies has been described as ``double asymptotic scaling". Roughly (but we will 
come to more precise statements below) Eq. (5) states that
$$F_2(x,Q^2)\simeq \exp\sqrt{|\log x|f_1}.\eqno (6)$$
On the other hand, and in the same notation,the findings of the 
HERA2 fits, in exponential form, imply
$$F_2(x,Q^2)\simeq \exp\{|\log x|f_2\},\eqno (7)$$
where the $f_i$ may depend weakly on $Q^2$. Undaunted by the evident incompatibility 
of (6) and (7), the authors in HERA2 happily assert at the same time the validity of 
(7), in their Table 4, and Eq. (8), and the validity of (6) in p. 26: ``Thus 
double asymptotic scaling is a dominant feature in this region".
 
\vskip.2cm
{\bf 4}.-How can these contradictory statements be reconciled? They cannot. The 
physicists in HERA2 have failed, in their theoretical analysis of the 
data, to do what an honest experimentalist should do, {\it viz.} to 
test theoretical assumptions without bias.

In fact there existed at the time of the first and 
(of course) the second HERA analyses {\it three}
 theoretical predictions (that I know of). From a standard moments analysis
 one can either have the ``double asymptotic scaling", refs.3, 5:
$$F_2(x,Q^2)\simeqsub_{x\rightarrow 0\atop Q^2\rightarrow \infty}
 C_0\left[ \frac{33-2n_f}{576\pi^2|\log x| \log[\alpha_s(Q_0^2)/\alpha_s (Q^2)]}
\right]^{\frac{1}{4}}
\exp \sqrt{\frac{144|\log x|}{(33-2n_f)}\;\left[\log\frac{\alpha_s(Q_0^2)}{\alpha_s(Q^2)}\right]}$$
$$+B_{NS}[\alpha_s(Q^2)]^{-D_{11}}x^{0.5}.
\eqno (8)$$ 
This depends on the singularity of $D(n)$ being the rightmost one. 
On the other hand, if it is the singularity of $a_n$ that lies to the right,
 one gets the factorization  behaviour given in ref.6:
$$F_2(x,Q^2)\simeq B_S[\alpha_s(Q^2)]^{-d_+}x^{-\lambda}
+B_{NS}[\alpha_s(Q^2)]^{-D_{11}}x^{0.5};\eqno (9)$$
I have added the nonsinglet contribution
 $B_{NS}[\alpha_s(Q^2)]^{-D_{11}}x^{0.5},\,D_{11}\simeq 0.512$, to both expressions for 
future ease of 
reference. In (9), $d_+$ is the largest eigenvalue of $D$, easily calculated; 
an explicit formula for it may be found in refs.4 or 6.
 It is to be stressed that, in 
the present state of 
the art, one cannot decide on the basis of perturbative QCD alone\footnote*{For 
deep inelastic $\gamma p$ scattering. For $\gamma^*\gamma$ the situation 
is different, and will 
be discussed below.} on which is 
valid, (8) or (9): it should be for experimentalists to tell us.

There is yet a {\it third} theoretical estimate$^{[7]}$. It is 
very difficult to compare with the other two, as it
 is obtained using totally different methods to get the prediction
$$F_2(x,Q^2)\simeq x^{-\omega_0\alpha_s}, \,\omega_0=\frac{4C_A\log 2}{\pi},\;C_A=3.\eqno (10)$$
Let me finish this point by telling what one gets if trying to fit (8), (9) or 
(10) to the HERA data. This test will be done 
in two steps. I will choose 
only data with $x<10^{-2}$, to be sure that 
one is in the ``small $x$" region and, in the first fits, I will also take
 $100\,\gev^2>Q^2_0>10\;\gev^2$ 
to avoid problems with varying number of excited flavours, large 
NLO corrections, and to have, for the test of (8), 
$Q^2\gg Q^2_0$, whatever the last may be. This region contains
 a total of 48 high-quality points. In the second step, the set of points 
will be enlarged to $Q^2$ up to $350\,\gev^2$. We will then have  
63 experimental points, distributed over a wide range in $Q^2$. Thus, the analysis 
will necessitate inclussion of NLO corrections. To the 
approximation needed, these may be implemented by simply 
taking $\alpha_s(Q^2)$ to two loops, and allowing $\Lambdav$ to 
vary freely, which is what I will do here.  

The results are now summarized, starting with the first situation.
 (10) runs contrary to the trend of the data, 
and produces a very large chi-squared, $\chi^2/{\rm d.o.f.}=926/(48-2)$. What is more, 
the fit rejects the introduction of a NS component, a clear indication of the 
fact that the behaviour (10) is certainly not attained at the HERA energies 
(the fit to the higher $Q^2$ data does not fare much better, either). Thus 
we will say no more about (10); the interested reader may find further discussion, 
and a possible re-interpretation of (10) in ref.4.

For the fits with Eqs.(8), (9), we fix $n_f=4$, take $\alpha_s$ to one loop,
$$\alpha_s(Q^2)=\frac{12\pi}{(33-2n_f)\log Q^2/\Lambdav},$$
and choose $\Lambdav=0.2\,\gev$ so that $\alpha_s(m_{\tau}^2)=0.32$. The results 
would not change much qualitatively if allowing, e.g., $\Lambdav$ to be 
a free parameter (some such results may be found in ref.4, and in 
the present note below).

The fit with the formula of De R\'ujula et al., {\it not} taking into 
account the NS piece is quite bad, but 
the results improve when including a NS contribution, as is done 
in Eq. (8). They are summarized 
in Table A1.
\vskip.2cm
\hrule
\vskip.1cm
$$\matrix{C_0&B_{NS}&Q_0^2&\chi^2/{\rm d.o.f.}\cr
7.1\times 10^{-3}&2.0&0.10\;\gev^2&51.3/(48-3)\cr}$$
\vskip.1cm
\centerline{{\bf Table } A1.- Fit with ``double scaling", Eq.(8). $\Lambdav=0.2$ (fixed)}
\vskip.1cm
\hrule\vskip.1cm\hrule
\vskip.2cm
This should, of course, be compared with the fit one gets (Table B1)
 if using the equation (9), 
since (9) and (8) stem from mutually incompatible assumptions. If we neglected 
the NS contribution in (9), we would also get a poor chi-squared; when including the NS piece 
this improves to about one unit by d.o.f., see Table B1.

\vskip.2cm
\hrule
\vskip.1cm
$$\matrix{B_S&B_{NS}&\lambda&\chi^2/{\rm d.o.f.}\cr
1.04\times 10^{-2}&1.21&0.40&49.8/(48-3)\cr}$$
\vskip.1cm
\centerline{{\bf Table } B1.- Fit with ``factorization", Eq.(9). $\Lambdav=0.2$ (fixed)}
\vskip.1cm
\hrule\vskip.1cm\hrule
\vskip.2cm
The results reported in Table B1 are more satisfactory than those given in 
Table A1: not so much the chi-squared by d.o.f., only marginally better, 
but because of the following other features. 
Firstly, the value of $B_{NS}$ in Table B1 is 
closer to what one obtains in the analysis of structure functions like $\nu W_3$ 
in neutrino scattering, which are pure 
nonsinglet and which suggest $B_{NS}\sim 0.6$. Another problem with the 
results reported in Table A1 is that they imply that the 
structure function at $Q_0^2$, essentially proportional to the 
Compton scattering cross section, should behave 
as a constant with the energy (proportional to $1/x$), which it does {\it not}: 
as shown in refs.4, 8, 9, it 
 still grows like a power of $1/x$.

 The situation is even more clear if we use the second set of HERA2 data with $x>10^{-2}$,
and all $Q^2$ between 12 and 350 $\gev^2$.  
Here one has to take into account NLO corrections, which, as explained, may be well 
approximated by just using the two-loop formula for $\alpha_s(Q^2)$,  
leaving $\Lambdav$ as a free parameter.

 For the ``double scaling" hypothesis, Eq.(8),  
the resulting chi-squared/d.o.f. {\it deteriorates}. 
Numerically, I have conducted a fit with (8) and $\alpha_s$ to two loops 
and find the results of Table A2.
\vskip.2cm
\hrule
\vskip.1cm
$$\matrix{\Lambdav&C_0&B_{NS}&Q_0^2&\chi^2/{\rm d.o.f.}\cr
0.40\,\gev&4.9\times 10^{-3}&2.0&0.38\;\gev^2&69.9/(63-4)\cr}$$
\vskip.1cm
\centerline{{\bf Table } A2.- Fit with ``double scaling",
 Eq.(8); $\Lambdav$ taken as a free parameter. $n_f=4$.}
\vskip.1cm
\hrule\vskip.1cm\hrule
\vskip.2cm
 That (8) does not give a very brilliant fit may also be 
 seen in Fig.10 of HERA2 (p.25), particularly in the lower part. 
In spite of the use of a log log plot, on 
which anything looks like a straight line, eyeball inspection of 
the scattering of the experimental points around the theoretical curve 
indicates a chi-squared/d.o.f. clearly larger than unity (the authors 
of HERA2 seem to be allergic to numbers, and do not give any figure for 
this or several other chi-squared of their fits). On top of it, the two problems 
mentioned in connection with the former fit, non-Pomeron behaviour 
at $Q^2_0=0.38\,\gev^2$ and deviation of $B_{NS}$ from its expected value 
persist.
 
For  Eq.(9), the details of the analysis 
may be found in ref.4; the chi-squared per d.o.f. 
actually {\it improves} to less than one unit, 
 and $B_{NS}$ decreases to the value of 0.8, quite close to the neutrino result of 0.6.
This is summarized in Table B2.
\vskip.2cm
\hrule
\vskip.1cm
$$\matrix{\Lambdav&B_0&B_{NS}&\lambda&\chi^2/{\rm d.o.f.}\cr
0.10\,\gev&1.0\times 10^{-3}&0.77&0.38&50.4/(63-4)\cr}$$
\vskip.1cm
\centerline{{\bf Table } B2.- Fit with ``factorization",
 Eq.(9); $\Lambdav$ taken as a free parameter. $n_f=4$.}
\vskip.1cm
\hrule\vskip.1cm\hrule
\vskip.2cm
The $\chi^2$/d.o.f. is now substantially better in the fit with (9) then in the 
fit with (8), 50.4 {vs.} 69.9; but other features also improve. Thus the value 
of $B_{NS}$ in Table B2 is now comfortably close to that expected from neutrino 
deep inelastic scattering, 0.77 compared to 0.6.

\vskip.2cm
{\bf 5.- Comments}.  One can of course still not not draw definite 
conclusions, but there is a clear trend: the data of HERA, both the old
 HERA1 and the new HERA2, support clearly a ``factorization" 
behaviour at small $x$, as predicted on the basis of the analysis of lower energy data 
long ago by L\'opez and the present author$^{[6]}$. As shown in the present note 
the ``double scaling"
 description, which the paper HERA2 claims to provide evidence for\footnote*{In all 
fairness it should be noted that the authors of HERA2 admit the 
existence of some deviations of the data from the predictions of double asymptotic scaling, 
e.g., in the last line of p.24.}  
(``double asymptotic scaling is a dominant feature") actually fails 
to give a good fit to the data, besides presenting extra 
problems. It is really sad that
 the HERA1 and HERA2 collaborations have not 
taken the trouble to present a fair analysis 
of experiment, not such a difficult task as I have been shown 
here. Why they ignore the possibility of a power-like behaviour as 
that given in Eq.(9) is really difficult to understand; apart from the 
original papers, it is given in the textbook of ref.10 considered (by e.g. the 
Particle Data Group) as one of the standard references in QCD. The situation 
is even more outrageous since many of
 the physicists involved in HERA1 and HERA2 have also been 
involved in the PETRA analyses of deep inelastic $\gamma^*\gamma$ scattering 
for which Witten$^{[11]}$ has derived (and experiment has checked) a behaviour, at 
small $x$, {\it exactly} like that in (9), with a value of $\lambda$,  $\lambda\sim0.5$ constant 
up to a slight dependence in the number of flavours excited. This was indeed 
one of the theoretical reasons given in the original papers$^{[6]}$ for 
preferring (9) to the ``double scaling" behaviour.    

\vfill\eject
{\bf REFERENCES}
\vskip.2cm

1.-M. Derrick et al, Z. Phys. {\bf C65} (1995) 397; Phys. Lett. {\bf B345} 
(1995) 576.\hb
2.-DESY 96-039, 1996\hb
3.-A. De R\'ujula et al.,  Phys. Rev{\bf D10} (1974) 1649.\hb
4.-F. J. Yndur\'ain, FTUAM 96-12 (hep-ph/9604263), 1996\hb
5.-S. Forte and R. D. Ball, CERN TH 95-323, 1995\hb
6.-C. L\'opez and F. J. Yndur\'ain, Nucl. Phys. {\bf B171} (1980) 231.\hb
7.-E. A. Kuraev, L. N. Lipatov and V. S. Fadin, Sov. Phys. JETP {\bf 44} (1976) 443; 
Ya. Ya. Balitskii and L. N. Lipatov, Sov. J. Nucl. Phys. {\bf 28} (1978) 822.\hb
8.-C. L\'opez and F. J. Yndur\'ain, Phys. Rev. Lett. {\bf 44} (1980), 1118.\hb
9.-M. Derrick et al, Phys Lett. {\bf B293} (1992) 465; Z. Phys. 
{\bf C63} (1995) 391.\hb
10.-F. J. Yndur\'ain, {\sl Quantum Chromodynamics}, Springer 1983; second 
edition as {\sl The Theory of Quark and Gluon Interactions}, Springer 1992.\hb
11.-E. Witten, Nucl. Phys. {\bf B120} (1977) 189.\hb
\vskip2cm
\noindent{\bf ACKNOWLEDGEMENTS}
\vskip.3cm
The author is grateful to F. Barreiro for very interesting discussions.
The financial support of CICYT, Spain, is also acknowledged.  
\end